\documentclass{sig-alternate}

\usepackage{amssymb}
\usepackage{url}
\usepackage{graphicx}
\usepackage[colorlinks=true,citecolor=blue,linkcolor=blue,urlcolor=blue]{hyperref}

\begin{document}
\conferenceinfo{IIWeb}{'12}
\title{Ranking Tweets Considering Trust and Relevance}

\numberofauthors{1}
\author{
\alignauthor
	Srijith Ravikumar,Raju Balakrishnan, and Subbarao Kambhampati \\
\affaddr{Computer Science and Engineering, Arizona State University} \\
\affaddr{Tempe AZ USA 85287}  \\
	\email{\{srijith,rajub,rao\}@asu.edu}
}

\maketitle

\begin{abstract}
The increasing popularity of Twitter and other microblogs makes improved trustworthiness and relevance assessment of microblogs evermore important. We propose a method of ranking of tweets considering trustworthiness and content based popularity.  The analysis of trustworthiness and popularity exploits the implicit relationships between the tweets.  We model microblog ecosystem as a  three-layer graph  consisting of : (i) users (ii) tweets and (iii) web pages. We propose to derive trust and popularity scores of entities in these three layers, and propagate the scores to tweets considering the inter-layer relations. Our preliminary evaluations show improvement in precision and trustworthiness over the baseline methods and acceptable computation timings.
\end{abstract}

\section{Introduction}

Twitter is increasingly used as a source of news and latest trends. Being open to all,  Twitter emerged as an excellent means to disseminate information to a large user community in the shortest time. On the negative side, this very open uncontrolled nature makes microblogging vulnerable to false information from malicious or credulous users~\cite{nytimestwitter,spamTwitterEconomist}. Recent trend of web search engines and online retailers considering the real-time trends in tweets for ranking products, news and recommendations aggravate this problem~\cite{abel2011analyzing,dong2010time} making microblog spamming more lucrative. Consequently, it is important to formulate sophisticated methods for analysis of relevance and trustworthiness for ranking tweets.

 Current Twitter ranking hearsay considers presence of query key words and recency of the tweets~\cite{rankingTwitter}. The increase in number of queries on a topic is generally associated with an increase in number of tweets. For example, when Apple releases a new model of iPhone, Twitter searches, as well as the tweets about the new model are likely to soar up.  Considering this correlation between tweets and searches, the popularity of a fact in tweets is a strong indicator of tweets' relevance. Twitter recognizes the importance of popularity, and assesses the popularity by the number of  retweets. While the number of retweets is an indication of popularity, this does not consider the content based popularity i.e. though two tweets are not retweets of each other, they may be semantically similar.  Secondly, considering the trustworthiness, retweeting need not indicate trust, as many users re-tweet without verifying the content. To get trustworthy tweets, Twitter tries to filter out spam tweets~\cite{spamTwitter}. While spam tweets are a form of untrustworthy tweets, providing correct information is more than just removing spam~\cite{nytimestwitter}. Even if the information is not deliberately manipulative, tweets may be incorrect.

To overcome these problems, we need a ranking sensitive to the content based popularity and trustworthiness of microblogs. Ranking should place the most credible and popular tweets in the top slots while searching with a keyword or \emph{hashtag}. To achieve this, we need methods to analyze the content based popularity and trustworthiness of individual tweets. Further, since the ranking is an online operation, the computational time should be acceptable. We believe that these problems are relevant not only to Twitter, but also to the search engines and retailers exploiting the Twitter trends for their rankings.

The main stumbling block in analyzing popularity and trustworthiness of tweets is that there is no authoritative source against which the information can be compared. Approaches like certifying  user profiles have limitations, since it is hard to verify millions of unknown and new users. Thus the very charm of open microblogging---\emph{anyone may say anything}---makes the problem harder. Further many users hardly verify the veracity of information before retweeting making propagation of false information easier. To deal with similar problems, web search engines use link analysis like PageRank~\cite{brin1998anatomy} to estimate the trustworthiness and importance of pages. Link analysis is not directly applicable to tweets since there are no hyperlinks between the tweets.

To surmount these hurdles, we propose to assess trustworthiness and popularity of tweets based on the analysis of the entire tweet ecosystem spanning across  tweets, users and the web. In the tweet space, we assess the  popularity of tweets based on the pair-wise content based agreement. On the web page space, we consider the page rank of the pages referred by the tweets. In the user space, we consider the implicit links between the users based on the follower-followee relationships. We propagate scores from all three layers based on the inter-layer relationships to compute a single tweet score.

We compare the credibility and relevance of the ranking by our method with the baselines. We show that the proposed method improves both the relevance and the trustworthiness of the top tweets compared to the baselines. Further timing experiments show that the computation time for the ranking is acceptable.

Rest of the paper is organized as the following. Next section describes the related work. The following section we present our model of the tweet space. Subsequently we describe our ranking methods, followed by section on experiments and results. Finally we present our conclusions and the planned future work.
\section{Related Work}
Ranking of tweets considering only relevance is researched extensively~\cite{TRECTwitter,duan2010empirical,nagmoti2010ranking}. Unlike our paper, these ranking approaches do not consider the trustworthiness.

Credibility analysis of Twitter stories have been attempted by Castillo \emph{et al.}~\cite{infocredibility}. The work tries to classify Twitter story threads as credible or incredible.  Our problem is different, since we try to assess the credibility of individual tweets. As the feature space is much smaller for an individual tweet---compared the Twitter story threads---the problem becomes harder.

Finding relevant and trustworthy results based on implicit and explicit network structures have been considered previously~\cite{gupta2011heterogeneous,sourcerank}. Real time web search considering tweet ranking has been attempted~\cite{abel2011analyzing,dong2010time}. We consider the inverse approach of considering the web page prestige to improve the ranking of the tweets. To the best of our knowledge, ranking of tweets considering trust and content popularity has not been attempted.
\section{ Modeling Twitter Ecosystem}
\label{sec:model}
\begin{figure}[t]
	\centering
	\includegraphics[scale= 0.5, trim=30mm 105mm 24mm 80mm, clip=true]{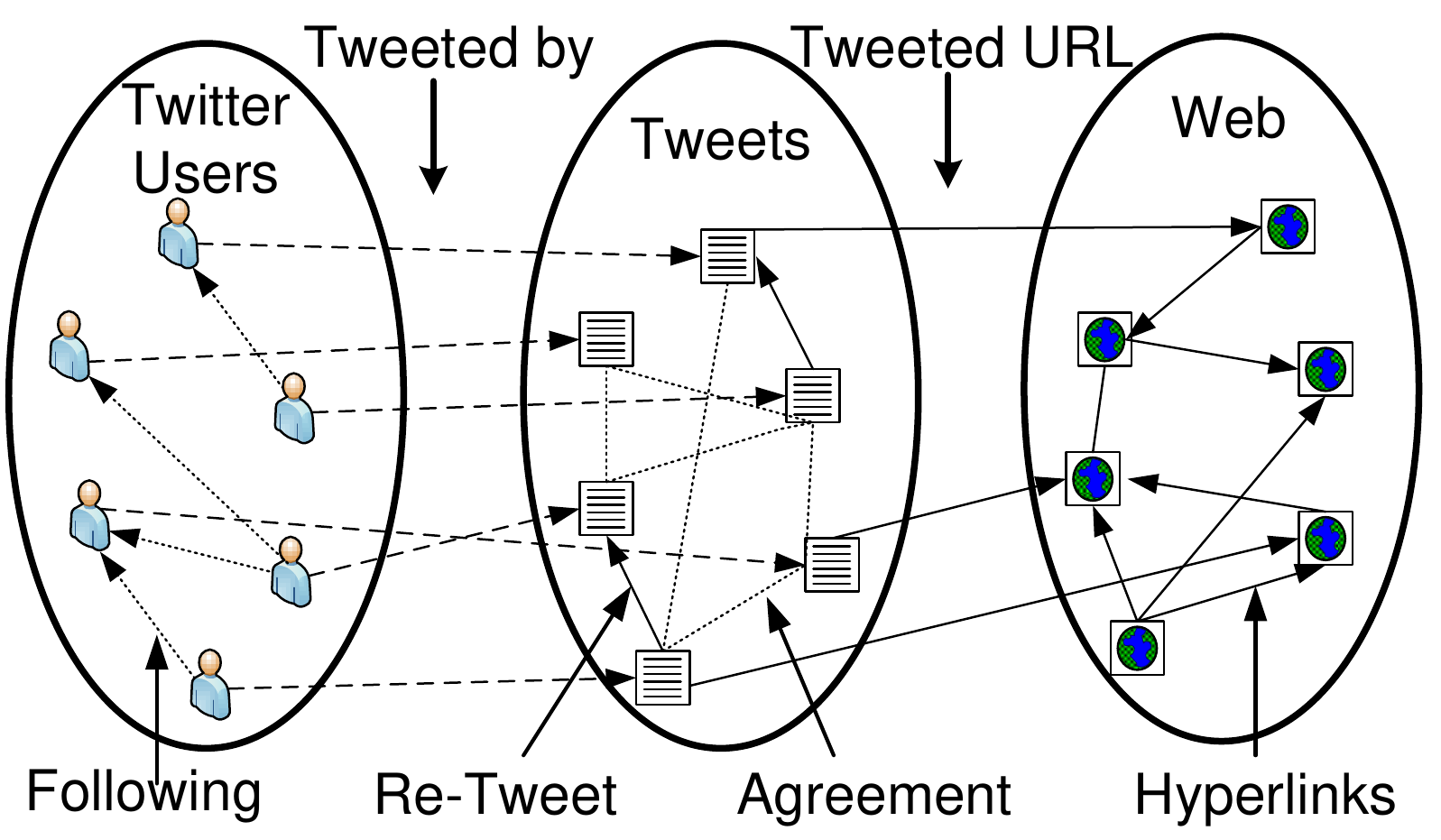}
	\caption{Three layer  ecosystem of Twitter space composed of user layer, tweets layer and the web layer. The  inter and intra layer edges are the implicit and explicit relations considered for the proposed ranking. }
	\label{fig:tweetModel}
\end{figure}

We model the entire tweet ecosystem as a three layer graph, as shown in  Figure~\ref{fig:tweetModel}. In the model the three layers are user layers composed of Twitter users, tweets layer composed of tweets and a web layer composed of pages. We exploit implicit and explicit links within the layers and across the layers for our ranking. The Twitter users are linked by \emph{who is following whom} relations. In the tweets layer, we build implicit links based on the content agreement, in addition to the directed retweet links. These agreement links provide evidence about many more tweets compared to very sparse retweet links. The web layer has explicit hyper links between pages. Though we considered only the relationships relevant to our ranking, other types of  relations may be derived in the space.

The proposed ranking is performed in the tweets layer. But we exploit all the three layers---user, web and tweets---to compute ranking scores. Within the tweets layer, we compute the content agreement between the tweets. Two tweets are in agreement if they have the same semantic sense. We will describe the details of the agreement computation in Section~\ref{subsec:agreementComputation}. In the user layer, we compute the scores of the users based on the following-followee relationships. These scores are propagated to the tweets by the \emph{Tweeted by} relationship. Similarly, we get the PageRank of the pages (which believed to be derived partially based on the hyperlinks in the web) referred by the tweets and propagate it back to derive ranking scores of the tweets.

\section{Ranking}
\label{sec:ranking}
In this paper we specifically focus on the ranking of tweets considering agreement. With respect to our model in Figure~\ref{fig:tweetModel}, this corresponds to ranking based on the agreement links in the tweets layer. The complete composite ranking exploiting all three layers are left for the future research.

\subsection{Agreement as a Basis of Ranking}
We explain the intuitions behind the agreement based ranking in this section. We compute the pair-wise agreement of tweets. A  tweet which is agreed upon by a large number of other  tweets is likely to be popular. Since popularity indicates relevance as we describe in the introduction, tweets with high agreement by other tweets are likely to be relevant. Alternatively, relevance assessment based on agreement may be viewed as an extension of relevance assessment exploiting the retweet based popularity.

With respect to the trustworthiness, if two independent tweeters agree on the same fact, tweets are likely to be trustworthy. The retweets are most likely not independent from the original tweets. Consequently, agreement is more indicative of trustworthiness than retweets. Please refer to Balakrishnan and Kambhampati~\cite{sourcerank} for a more general explanation of why agreement is likely to indicate trustworthiness and relevance.

\begin{figure*}[t]
	\begin{minipage}[b]{.47\textwidth}
	\includegraphics[scale=.4,trim= 100 70 100 80,clip=true]{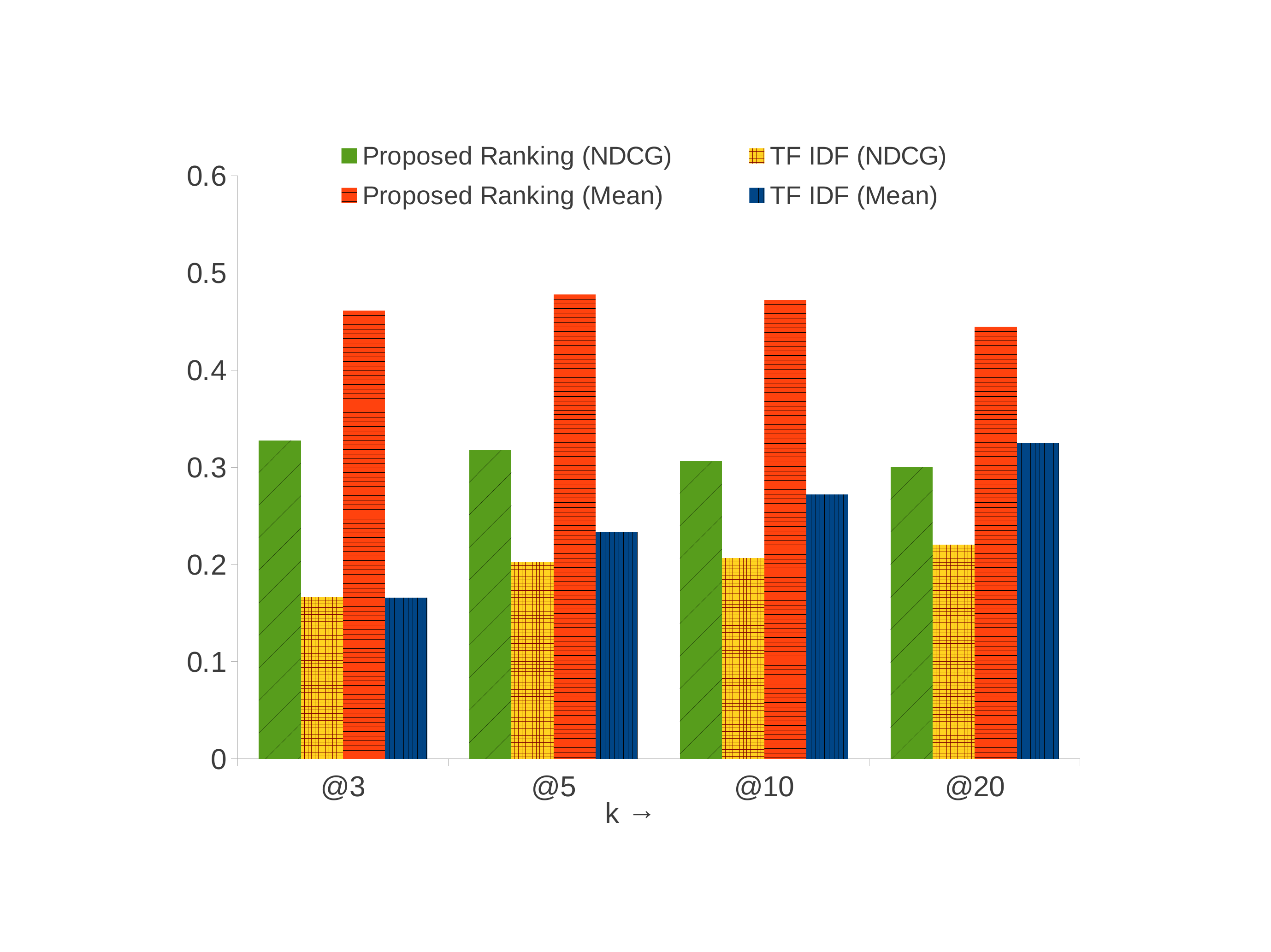}
	\caption{Top-K Results vs Relevance Measure}
	\label{fig:Relevance Measurement}
	\end{minipage}\qquad
	\begin{minipage}[b]{.47\textwidth}
	\includegraphics[scale=.4,trim= 100 70 100 80,clip=true ]{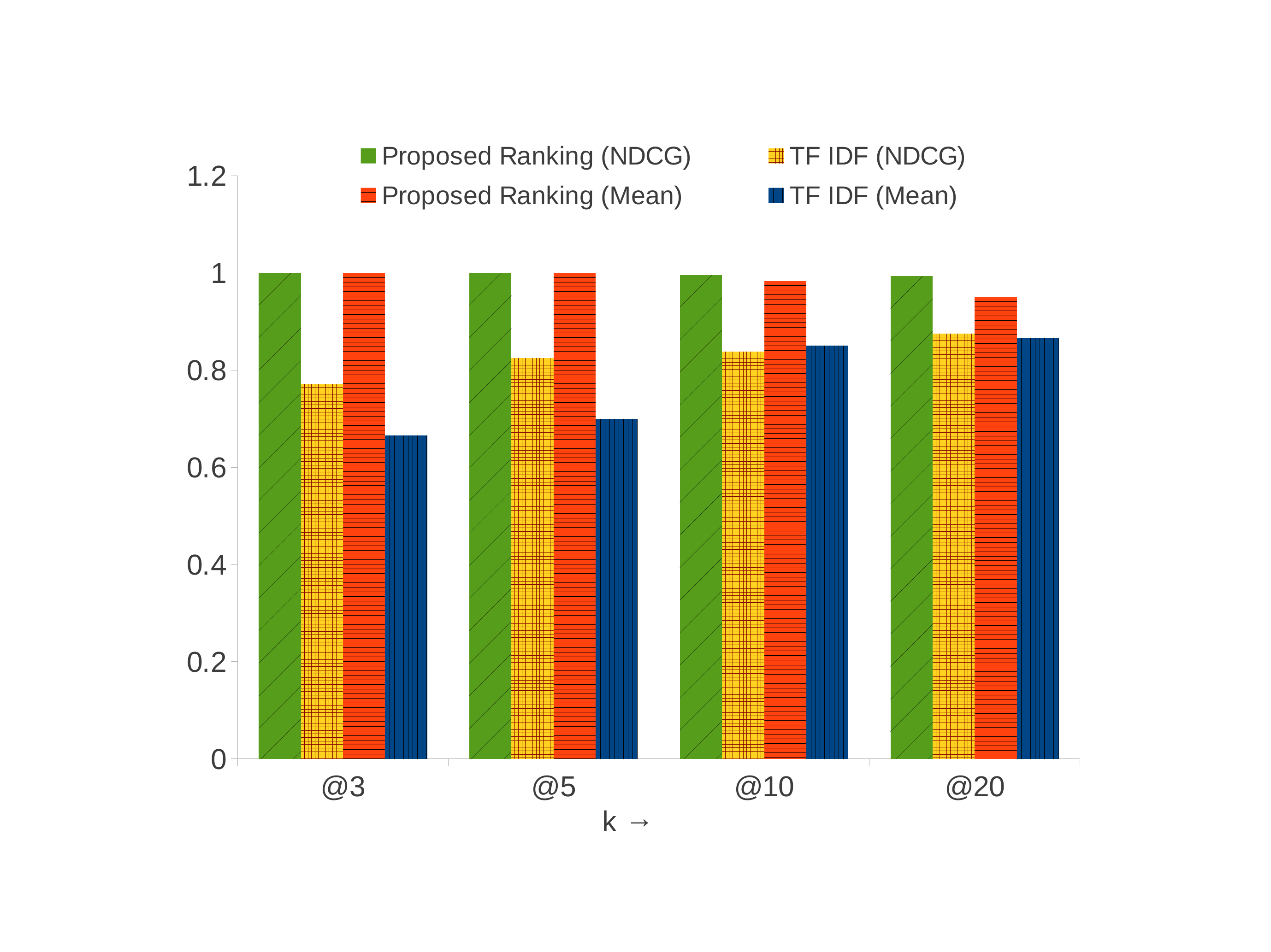}
	\caption{Top-K Results vs Trust Measure}
	\label{fig:Trust Measurement}
	\end{minipage}
\end{figure*}

\subsection{Agreement Computation}
\label{subsec:agreementComputation}

Computing semantic agreement between the tweets which satisfies the query-time constraints is challenging. We compute the agreement between the query based on  Soft-TFIDF, and calculated the ranking scores based on  voting.

Soft-TFIDF is similar to the normal TFIDF, but considers similar tokens in two compared document vectors in addition to the exactly same tokens. We use Soft-TFIDF with Jaro-Winkler similarity; which is found to perform well for named entity matching~\cite{cohen2003comparison} and computing semantic similarity between the web database entities~\cite{sourcerank}.

 Let $\mathcal{C}(\theta, v_i, v_j)$ be the set of words for
$w \in v_i$ such that there is some $u \in v_j$ with $sim(w,u) >
\theta$.  Let $D(w,v_j)=max_{u\in v_j}sim(w,u)$. The
$\mathcal{V}(w,v_i)$ are the normal TF values weighted by $log(IDF)$
used in the basic TF-IDF. SoftTFIDF is calculated as,
\begin{equation}\mathcal{SIM}(v_i,v_j)=\sum_{w\in \mathcal{C}(\theta,v_i ,v_j)}\mathcal{V}(w,v_i)\mathcal{V}(u,v_j)D(w,v_j)\end{equation}
We used Jaro-Winkler as the secondary distance fucntion for the  $sim$ function above. Parameter $\theta$ is set to $0.6$, as this value was found to be performing well based on cross-validation.

To formulate the final ranking combining agreement, keyword based similarity and recency of tweets, we send queries to Twitter and retrieve top-$N$ (we used $N=200$) tweets. After computing the pair-wise similarity between the tweets as described above, we represent the tweets as weighted graph with tweets as vertices and edges as similarity (this graph based representation makes some of our future research easier) In this weighted graph, we compute the score for a tweet as the sum of its' the edge weights. Finally we rank the tweets based on this edge weight score and present the top-$k$ to the user. Since the top-$N$ tweets are returned by Twitter considering keyword relevance and recency of the tweets, these two factors are implicitly accounted in the proposed ranking.

\section{Evaluation}
We conducted a preliminary evaluation of the proposed ranking method against popular ranking of TF-IDF based on query similarity. We compared the top-$k$ precision and Normalized Discounted Cumulative Gain (NDCG) of the proposed method with the TF-IDF. Subsequently we compared trustworthiness of the top-$k$ tweets by the proposed method with the baselines. Further, we evaluated the variation of computation timings with the size of the ranked tweet set.
\subsection{Test Tweet Set}

We used the Twitter's trending topics spanning across current news, sports and celebrity gossips for our evaluations. Trending topics are used to get enough number of tweets with varying degrees of trustworthiness and relevance. For each topic,  top 1500 tweets are retrieved using the Twitter API (1500 is the maximum number of tweets returned by the Twitter API). The tweets marked as retweets  are removed. We randomly sampled 200 tweets from these 1500 tweets to get our test set. We used a random sample of 200 tweets instead of top-200 results from Twitter, as often the top-k tweets contains repetitions of a few tweets; since many users copy-past same information without explicitly retweeting. Thus randomly sampled 200 tweets from top 1500 tweets increases the variance in the tweet quality in the test set so that different ranking methods can be better distinguished. We used enough number of queries to distinguish the proposed method from the TF-IDF with a statistical significance of 0.8 or above in every experiment below.\footnote{We will improve the significance level to 0.9  in our future experiments.}

\subsection{Relevance Evaluations}
To assess the relevance, we manually labeled the tweets  with a relevance value of 0, $\frac{1}{3}$, $\frac{2}{3}$ and 1. The test data for 6 search queries contained 187 tweets of zero relevance to the query, 473 tweets of relevance $\frac{1}{3}$, 249 tweets of relevance $\frac{2}{3}$ and 39 tweets of relevance 1. The classification was done based on the relevance of the tweet to the current news matching that trending topic. For example, if the topic is ``\textit{britney spears}" and the current news during the tweet generation were about Britney Spears engagement, the tweets which were not related to the trending topic or spam are given a score of 0 (e.g. \textit{I liked a @YouTube video Britney Spears}),  the tweets which are remotely relevant were given a score of $\frac{1}{3}$ (e.g. \textit{Britney Spears Is Engaged}), tweets which have some information on engagement were given a score of $\frac{2}{3}$ (e.g. \textit{Britney Spears engaged to marry Jason Trawick (AP)}), and the tweets which have good amount of information are given a perfect score of 1 (e.g. \textit{@BritneySpears engaged to marry her longtime boyfriend and former agent Jason Trawick}).

The comparison of top-$k$ precision of the proposed method with the TF-IDF is shown in Figure~\ref{fig:Relevance Measurement}. The proposed method improves both NDCG and top-$k$ precision for all values of $k$. Note that the apparently low value of mean relevance (less than 0.5) is due to the fact that only a very small fraction of tweets have high relevance values. Though a direct comparison is not possible with TREC 2011 microblog track results as the data is not publicly available yet,  top precisions in TREC  are in comparable ranges~\cite{TRECTwitter}.

\subsection{Trust Evaluations}
Similar to the relevance evaluations, we labeled the tweets as trustworthy or untrustworthy manually. Tweets were given a scores of -1, 0 or 1, where -1 is for the untrustworthy tweets such as spam or are wrong facts (e.g.\textit{Britney Spears engaged to a Sachem alum.}), 0 for tweets which are opinions (e.g. \textit{We can all rest now \#Britney}) and 1 to the tweets which contain correct facts (e.g. \textit{Britney Spears is engaged to marry Jason Trawick}). Our dataset for the 6 queries contained 29 tweets with score -1, 157 tweets of score 0 and  742 tweets of score 1. Note that the returned tweets are after the spam filtering by the Twitter which itself eliminates many spam tweets.

Figure \ref{fig:Trust Measurement} shows the comparison of the proposed method with TF-IDF based ranking. The top-$k$ tweets returned by the proposed method are almost always trustworthy, whereas the TF-IDF returns many of the untrustworthy tweets in the top. This shows that the proposed method effectively removes the untrustworthy tweets and returns trustworthy ones in the top slots, even for $k=20$.
\begin{figure}[t]
	\centering
	\includegraphics[scale=0.27,trim= 65 34 49 16,clip=true ]{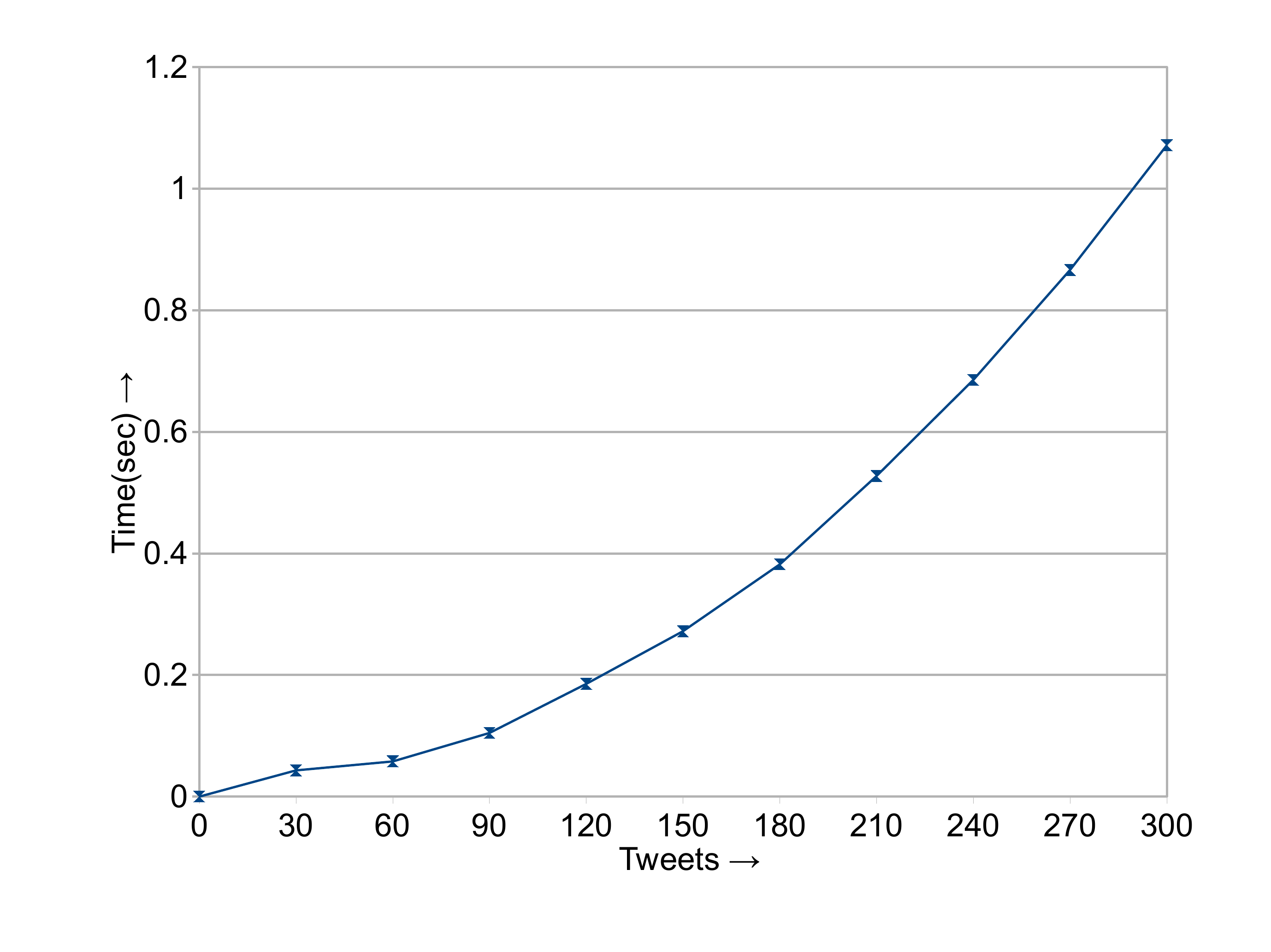}
	\caption{ Number of Tweets vs computation time.}
	\label{fig:Time Measurement}
\end{figure}
\subsection{Timing Evaluation}
As the ranking is at the query time, computation time must be within acceptable limits. We evaluated the time taken for ranking against the number of ranked tweets. The experiments are performed in a dual core 3 GHz machine with memory of 8 GB. In  Figure~\ref{fig:Time Measurement}, ranking up to 300 tweets takes less than 1.2 seconds. The proposed approach of selecting top tweets based on the recency and further ranking the selected set of tweets, of the order of hundreds, is feasible (note that our experiments used only 200 tweets). The time increases quadratically in the number of tweets as expected. Further, notice that computation of the pairwise agreement---the time consuming part of the ranking---can be easily parallelized (e.g. using MapReduce) since agreement computation can be performed in isolation without interprocess computation.

\section{Conclusions and Future Work}
In order to rank the tweets, consideration of content based popularity and trustworthiness is essential. Towards this end, we model the Twitter ecosystem as a tri-layer---user, tweets and web layers---graph and propose a ranking exploiting explicit and implicit links in the three layers. As the first step towards a complete ranking, we formulate a ranking based on agreement of tweets. Our initial evaluations show improvement of precision and trustworthiness by the proposed ranking and acceptable computation timings.

We plan to extend the method in several directions. In addition to the currently considered agreement, recency and keyword similarity, we propose to exploit web and user layers  to formulate a composite ranking. In the user layer, we plan to consider the credibility of the users based on the follower relationships and past tweets. Subsequently, author credibility will be propagated to the tweets for ranking. In the web layer, we plan to consider the reputation of the pages referred by the tweets. Further, we plan to have enhanced agreement computations and extensive user evaluations.

\bibliographystyle{abbrv}
\bibliography{microblog}
\end{document}